\documentclass[prb,twocolumn,showpacs,floatfix,superscriptaddress]{revtex4}
\usepackage{graphicx,graphics,color,epsfig}
\usepackage{bm}
\usepackage{amsmath}
\usepackage{dsfont}
\usepackage{amsfonts}
\usepackage{amssymb}
\usepackage{epstopdf}

\begin{document}

\title{3D Non-Abelian Anyons: Degeneracy Splitting and Detection by Adiabatic Cooling}

\author{Seiji J. Yamamoto}
\affiliation{National High Magnetic Field Laboratory and Department of Physics, Florida State University, Tallahassee, Florida 32306, USA}

\author{Michael Freedman}
\affiliation{Microsoft Research, Station Q, Elings Hall, University of California, Santa Barbara, CA 93106, USA}

\author{Kun Yang}
\affiliation{National High Magnetic Field Laboratory and Department of Physics, Florida State University, Tallahassee, Florida 32306, USA}

\date{\today}

\begin{abstract}
Three-dimensional non-Abelian anyons have been theoretically proposed to exist in heterostructures
composed of type II superconductors and topological insulators.
We use realistic material parameters for a device derived from Bi$_2$Se$_3$
to quantitatively predict the temperature and magnetic field regimes where an experiment
might detect the presence of these exotic states by means of a cooling effect.
Within the appropriate parameter regime, an adiabatic increase of the magnetic field
will result in a decrease of system temperature when anyons are present.
If anyons are not present, the same experiment will result in heating.
\end{abstract}


\maketitle

\section{Introduction}

Majorana fermions (which are responsible for the non-Abelian content of Ising anyons \cite{Nayak2008})
have been proposed to exist in a number of condensed matter settings.
One particularly intriguing development involves the Majorana states trapped within the hedgehog defects that
emerge at the interface between a topological insulator and the mouths of vortices in a type II superconductor \cite{Teo2010}.
Since the realization of anyon statistics is generally believed to be impossible beyond two dimensions,
the novelty of this setup is its three dimensional (3D) nature.
The paradox resolves upon realizing that the anyons at hedgehogs
are not merely point-like objects, but
singularities of a $U(n)/O(n)$ order parameter which may be modeled
by ribbons joining hedgehogs \cite{Freedman2011}.
The ribbons contain additional twist degrees of freedom relevant to
particle exchange:
twisting a ribbon transforms the state.
The anyons in the Teo-Kane model provide a projective
representation of the ribbon permutation group \cite{Freedman2011}.
Interestingly, this remnant of braiding in higher dimensions seems to be special to Ising anyons:
Fibonacci anyons, for example, seem to have no 3D version.
While these mathematical details are interesting, they are subordinate to the question relevant to physicists:
Do 3D anyons really exist or are they merely theoretical constructs?
To the best of our knowledge, this paper provides the first proposal to settle this
question in the lab rather than via theoretical arguments.

Majorana fermions have proven challenging to unambiguously detect in any dimension 
not to mention the exotic 3D context we have in mind here.
For example, within the setting of quantum Hall systems, relevant to 2D non-Abelian anyons,
experiments usually involve edge state transport \cite{Nayak2008}.
Recently, however, several \textit{bulk} probes have been proposed
\cite{Yang2009, Cooper2009}.
In particular, one suggestion exploits the huge ground state degeneracy inherent to the non-Abelian
anyon system to produce a cooling effect \cite{Gervais2010}.
This can be understood in analogy to adiabatic cooling via spin demagnetization,
but here the entropy reservoir is an anyon system rather than a spin system.
Crucially, the non-Abelian entropy is temperature independent but proportional
to the number of non-Abelian anyons, which,
in turn, can be adiabatically (or, more precisely, isentropically) controlled by a magnetic field.
Thus, an adiabatic increase in the non-Abelian entropy necessarily implies a decrease in the rest of the system's
entropy. This is accomplished by a lowering of the system's temperature.  While this idea has been explored in a 2D
setting \cite{Gervais2010}, we expect it to work even better in 3D where the
topological and conventional sources of entropy (which are fundamentally 3D)
will be on an equal footing.

In this paper we explore the possibility of detecting 3D anyons via cooling by applying the Teo-Kane model to
a specific heterostructure involving a topological insulator and a superconductor.  Since we are mainly interested
in demonstrating feasibility, our principal goal will be to establish an experimentally accessible temperature window
$T_{L} < T < T_{U}$
in which the cooling effect is significant.
We make quantitative estimates using realistic material parameters.
The upper bound of the temperature window, $T_{U}$, is determined by the temperature where sources of entropy
other than anyons begin to dominate the system.
The lower bound of the temperature window, $T_{L}$, is established by calculating
the ground state energy degeneracy splitting
that results from Majorana-Majorana tunneling.  Operating at temperatures above
this splitting energy allows us to treat the low energy Hilbert space as essentially degenerate.

In Section \ref{sec:MajoranaDetection1d2d} we briefly mention existing proposals to detect the more conventional
type of non-abelian Ising anyons below three dimensions.  Section \ref{sec:DevStruc} describes the physical
structure and material components of the device for our 3D anyon system.  
The model we use to analyze this device is explained in Section \ref{sec:Model},
from which we construct Majorana solutions in Section \ref{sec:MajoranaSols} and
calculate the ground state degeneracy splitting as a function of Majorana separation in Section \ref{sec:Degen}.
Section \ref{sec:Entropy} describes the calculation of all contributions to the system entropy.
We use this in Section \ref{sec:Cooling} to determine the feasibility of detection by cooling
and establish the parameter regime in which to carry out the search for 3D non-abelian anyons.
Our conclusions are summarized in \ref{sec:Conclusions}.

\section{Majorana Detection Proposals for $D<3$}
\label{sec:MajoranaDetection1d2d}
There are now many proposals for condensed matter realizations of Majoranas in two dimensional systems.
Among them, the most actively studied thus far is the quantum Hall state at filling factor $5/2$ where 
detection schemes have been proposed using both edge \cite{Stern2006, Bonderson2006} and bulk \cite{Yang2009, Cooper2009} probes. 
Some tantalizing evidence has been reported based on the former \cite{Willett2009, Willett2010}.
Other systems in which Majoranas may exist and detection strategies have been theoretically proposed include
strontium ruthenates \cite{DasSarma2006},
helium-3 \cite{Tsutsumi2008},
topological insulator based heterostructures \cite{Fu2009},
semiconductor heterostructures \cite{Sau2010, Alicea2010},
cold atoms systems \cite{Zhu2011},
and
quantum wire networks \cite{Kitaev2001, Oreg2010, Lutchyn2010, Alicea2011}.
The experimental confirmation of the existence of these 2D Majoranas has lagged somewhat the large
number of theoretical propositions.

Nonetheless, all these Majorana fermions are of the two dimensional SU(2)$_2$ Ising anyon variety
\cite{Nayak2008}.
They are representations of the braid group.
The possibility for anyons beyond two dimensions came as a bit of a shock \cite{Nayak2010, Stern2010}
due to certain long-standing trusted mathematical arguments.
It turns out that no mathematical theorems need to be corrected because these 3D anyons do not provide representations of the braid group but rather provide projective representations of the ribbon permutation group \cite{Freedman2011}.

We also mention in passing that the quantum wire networks involve 1D Majorana fermions in a sense, but their braiding (for example using T-junctions as proposed by Alicea et al \cite{Alicea2011}) still requires the spanning of 2D real space by the network.  Furthermore, these 1D Majoranas are still linked to the 2D anyon statistics tied up with the braid group rather than the 3D projective ribbon permutation statistics of 3D anyons.

\section{Device Structure and Materials}
\label{sec:DevStruc}
\begin{figure}[tbp]
   \centering
   \includegraphics[width=3.4in]{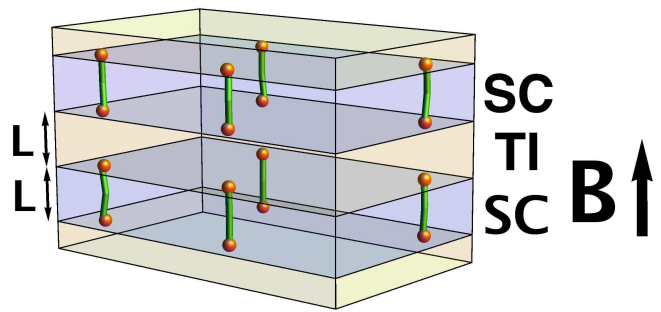}
   \caption{
   (color online) Two superlayers of the heterostructure consisting of alternating layers of topological insulator (TI)
   and superconductor (SC).  The external magnetic field induces vortices (green tubes) in the SC layers
   that terminate at the interface with the TI layers resulting in 3D anyons (orange balls).
   In addition to the obvious three dimensional distribution of anyons in real space, these anyons
   provide a projective representation of the ribbon permutation group~\cite{Freedman2011} which further distinguishes
   them from the more well-known two dimensional anyon system (which forms a representation of the braid group).
   }
   \label{fig:figure1}
\end{figure}

Consider an s-wave, type II superconductor sandwiched between layers of a 3D topological insulator.
The superconductor occupies the region $-\frac{L}{2} < z< \frac{L}{2}$
with flanking layers of topological insulator for $\frac{L}{2}< |z| < L$.
This structure is repeated along the $\hat{z}$-direction to
build a superlattice and thus a true bulk 3D system.  See figure~\ref{fig:figure1}.
Due to this periodicity, our calculations will only need to consider a single superlayer.
For the topological insulator, we choose Bi$_2$Se$_3$ appropriately p-doped so that the Fermi level
($\epsilon_F$) is pushed down into the bulk gap.
For the superconductor, we choose the n-doped material Cu$_{0.12}$Bi$_2$Se$_3$
which is known to be a strongly type II ($\kappa = \lambda/\xi \approx 50$) bulk superconductor \cite{Hor2010}.
The precise nature of the superconductivity in this material is not yet known,
but a theoretical proposal has suggested that it may be unconventional \cite{Fu2010}. In the absence of further
experimental data, we will assume it to be s-wave; this and other caveats are discussed further at the end of the paper.

Existing experimental results on Cu$_{0.12}$Bi$_2$Se$_3$ (almost) supply the required minimal input needed for our calculation.
ARPES provides numbers for the Fermi energy, velocity, wavevector, and the spin-orbit gap \cite{Wray2010}:
$\epsilon_F = 400 \text{ meV}$,
$\hbar v_F = 3.8 \text{ eV \AA}$,
$k_F = 0.1 \text{ \AA}^{-1} \approx \epsilon_F/\hbar v_F$, and
$\Delta_{so} = 150 \text{ meV}$.
We also need the superconducting gap ($\Delta$) and coherence length ($\xi$).
Based on the existing experimental data, there are two ways of determining $\Delta$ and $\xi$.
Both quantities could be inferred from the known superconducting transition temperature, $T_c = 3.8$ K
\cite{Hor2010}, using the BCS relations $\Delta = 1.764k_B T_c$ and $\xi = \hbar v_F/\pi\Delta$.
On the other hand, we could use the measurement of the upper critical field, $B_{c2} = 1.7$ Tesla \cite{Hor2010},
to give a more direct estimate of the coherence length $\xi = \sqrt{\frac{h/2e}{2\pi B_{c2}}} \approx 139$ \AA\,
independent of BCS theory.
This could then be combined with the BCS relation $\Delta = \hbar v_F/\pi\xi$ to give a value for the superconducting gap.
The fact that these two methods do not agree \cite{Hor2010,Wray2010} provides a puzzle for the community.
What is needed is a \textit{direct} measurement of $\Delta$,
which has not been reported yet.
Since this paper is chiefly concerned with vortices we will trust the $B_{c2}$-derived value of $\xi$, which is $139$ \AA.
To be consistent, this value of $\xi$ is then used to set $\Delta = 8.7$ meV.

Very recently, another indirect estimate of the gap has been made using specific heat data
\cite{Kriener2011}.
Weak-coupling BCS theory does not fit the data, while a fit to strong-coupling BCS theory produces parameter values
inconsistent with other measures.  Therefore, the main conclusions to be drawn from this experiment are 
the bulk nature of the superconductivity,
the lack of nodes in the gap, 
and the fact that the gap is larger than would be expected from BCS theory using $T_c$.  
We therefore continue to use $B_{c2}$ to estimate the value of the superconducting gap.

\section{Theoretical Model}
\label{sec:Model}
The low-energy theory is an eight-band Dirac model
\cite{Teo2010, Nishida2010, Fukui2010}:
$H = \frac{1}{2} \int d^3x \Psi^{\dagger} \mathcal{H} \Psi $ where
\begin{eqnarray}
	\mathcal{H}
	&=&
\left(
\begin{array}{cc}
	\mathcal{H}_D 	& \Delta \\
  	\Delta^*		&  -\mathcal{H}_D
\end{array}
\right)
\end{eqnarray}
with diagonal terms given by
$\mathcal{H}_D =  \bm{\alpha}\cdot \mathbf{p} -\epsilon_F - i \gamma^5 \beta \Delta_{so}$.
We use the standard Dirac-Pauli representation:
$
\bm{\alpha} = \left( \begin{array}{cc}
		0 & \bm{\sigma} \\
		 \bm{\sigma} & 0
	\end{array}\right)
$,
$
\beta = \left( \begin{array}{cc}
		\mathds{1} & 0 \\
		0 & -\mathds{1}
	\end{array}\right)
$,
$
\gamma^5 = \left( \begin{array}{cc}
		0 & \mathds{1} \\
		\mathds{1} & 0
	\end{array}\right)
$.
The fermion is an eight-component object deriving from spin, orbital, and particle-hole degrees of freedom:
$\Psi^{\dagger} = \left( \psi^{\dagger}, -i \psi^{T} \alpha_2 \right) = (u_1^{\dagger}, u_2^{\dagger}, v_2^{\dagger}, v_1^{\dagger})$,
where each $u_{i}$ and $v_{i}$ has two components.
Particle-hole symmetry enforces $v_{i} = i\sigma_2 u_{i}^*$.
We have written the model in the same way as Ref.~\onlinecite{Nishida2010},
which is unitarily equivalent to the models used in Refs.~\onlinecite{Teo2010} and \onlinecite{Fukui2010}.
Variants of this model have been studied for many years \cite{Jackiw1976}.

We use a Dirac Bogoliubov-de-Gennes (BdG) model for this material.  The construction
of the effective model for Bi$_2$Se$_3$ has been discussed by several groups, see 
for example \cite{Liu2010} and references therein.  While higher momentum terms
can sometimes lead to interesting physics, the situation we are considering only requires
use of a Dirac model.  

\section{Majorana Solutions}
\label{sec:MajoranaSols}
We combine the superconducting and spin-orbit gaps into a single 3-component
order parameter \cite{Teo2010}: $\vec{n} = (\text{Re} \Delta, \text{Im} \Delta, \Delta_{so})$.
Within the superconductor, a vortex excitation along the $z$-axis is achieved by imposing
the following profile on the superconducting order parameter:
$
	\Delta(\rho,\theta,z) = \Delta(\rho) e^{in\theta} \Theta(L/2 - |z|).
$
The interface between the topological insulator and superconductor is specified
by imposing the following kink profile on the spin-orbit gap:
$\Delta_{so} (\rho,\theta,z) = +\Delta_{so}$ for $|z| \ll L/2$
and
$\Delta_{so} (\rho,\theta,z) = -\Delta_{so}$ for $|z| \gg L/2$.
In this way the band is inverted within the topological insulator, while taking the opposite sign
in the (topologically trivial) superconductor.
Note that we have considered a topological insulator and trivial superconductor,
but the kink would also exist at the interface of a trivial insulator and a topological superconductor.

The combination of a vortex in $\Delta$ and a $Z_2$ kink in $\Delta_{so}$ leads to an anisotropic hedgehog
in $\vec{n}$ occurring where the vortex tube meets the interface with the topological insulator.
We can think of this defect as a potential well and solve for the zero energy solutions of the BdG equation
$\mathcal{H} \Psi = 0$.  At the kink ($z=-L/2$) this leads to the following Majorana zero-mode wavefunctions
\cite{Nishida2010, Fukui2010}
\begin{eqnarray}
	\left( \begin{array}{c}
		u_{1\uparrow} \\
		u_{1\downarrow} \\
		u_{2\uparrow} \\
		u_{2\downarrow}
	\end{array}\right)
	&=&
	\mathcal{N}
	\left( \begin{array}{c}
		J_{(n-1)/2}\left( k_F \rho \right) e^{-i\pi/4}e^{i(n-1)\theta/2} \\
		0\\
		0\\
		J_{(n+1)/2}\left( k_F \rho \right) e^{i\pi/4}e^{i(n+1)\theta/2}
	\end{array}\right) \nonumber\\
	&&\times	
		\sqrt{ \epsilon_F }
		e^{-\frac{1}{\hbar v_F} \int^{\rho} \Delta(\rho^{\prime})d\rho^{\prime}}
		e^{-\frac{1}{\hbar v_F}\int^{z} \Delta_{so}(z^{\prime})dz^{\prime}} \quad
\label{eq:MajoranaKink}
\end{eqnarray}
while at the upper interface ($z=L/2$) we have an anti-kink
given by cyclically permuting the right hand side two steps.
The expressions for $v_{i}$ follow from particle-hole symmetry
and $\mathcal{N}$ is a normalization constant given by
$\mathcal{N}^{2} = \frac{\epsilon_F}{2\hbar^2v_F^2  \xi_z[E(-\epsilon_F^2/\Delta^2) - K(-\epsilon_F^2/\Delta^2)]}$
where $K$ and $E$ are the complete elliptic integrals of the first and second kinds.
These expressions are valid inside the superconductor.
We ignore the exponential tail decaying into the topological insulator
which is miniscule due to the large insulating gap.

An exact treatment of this problem would numerically solve for the order parameter profiles self-consistently.
See, for example, Ref.~\onlinecite{Gygi1991}.
However, since we desire analytical expressions we make the following
standard simplification: $
e^{-\frac{1}{\hbar v_F} \int^{\rho} \Delta(\rho^{\prime})d\rho^{\prime}}
e^{-\frac{1}{\hbar v_F}\int^{z} \Delta_{so}(z^{\prime})dz^{\prime}}
=
\text{sech}(\rho/\pi\xi) \text{sech} (z/\xi_z)$
where $\rho \equiv \sqrt{x^2+y^2}$ and $\xi_z \equiv \hbar v_F/\Delta_{so} \approx 25.3$ \AA .
Thus, two experimentally determined parameters influence the localization of the Majorana wavefunction:
$\xi$ and $\Delta_{so}$ (or, equivalently, $\xi_z$ as defined above).
$\xi_{z}$  determines the localization in the $z$-direction
while 
$\xi$ sets the decay length in the $xy$-plane.
Since $\xi \gg \xi_z$, we might already speculate that the in-plane
coupling between Majoranas will be much more important than the
coupling in the $z$-direction for the degeneracy splitting.
We turn to this issue next.

\section{Degeneracy Splitting}
\label{sec:Degen}
With the Majorana wavefunctions in hand, we can use these expressions to calculate the
splitting of the ground state degeneracy as a function of Majorana separation.
This energy splitting has been calculated by a variety of means in 2D Majorana systems.
We will generalize to 3D the method of Ref.~\onlinecite{Cheng2010}
who adapted to 2D the 1D Lifshitz problem \cite{Landau1974}.
Our calculation is very similar to what has already been presented in Refs.~\onlinecite{Cheng2010}
and \onlinecite{Mizushima2010}.  We refer the reader to these papers for technical details.
In what follows, we outline the main idea behind the calculation.

Consider two Majorana states, $\Psi_{a}$ and $\Psi_{b}$, which are brought together from infinity.
As they approach each other, the degenerate eigenvalue of the two fusion
channels is split by an amount $E_{\text{split}}$.  The new eigenfunctions of this two-Majorana state are
$\Psi_{\pm} \equiv \Psi_a \pm i \Psi_b$ with corresponding eigenvalues $E_{\pm}$.
Particle-hole symmetry dictates $E_{\text{split}} = E_{+} - E_{-} = 2E_{+}$.
We calculate the eigenvalue of one of these states as follows \cite{Cheng2010}:
\begin{eqnarray}
	E_{+} &=&  \frac{\int_{\Sigma} \Psi^{\dagger}_a \mathcal{H} \Psi_{+} - \int_{\Sigma} \Psi^{\dagger}_{+} \mathcal{H} \Psi_{a} }{\int_{\Sigma} \Psi^{\dagger}_a \Psi_{+} }
	\label{eq:Esplit}
\end{eqnarray}
where $\Sigma$ is an integration region corresponding to the half-infinite volume in three dimensions.
The integrand can be written in terms of total derivatives, and the localized nature of the wavefunctions
allows us to reduce $\Sigma$ to the infinite 2d plane that bisects the line joining the two Majorana states in question.
We consider two cases:
vertical Majorana-Majorana coupling in the $z$-direction and
lateral Majorana-Majorana coupling in the $xy$-plane.

For the vertical coupling, we place
$\Psi_{a}$ at $\mathbf{R}_{a} = (0,0,-L/2)$ and
$\Psi_{b}$ at $\mathbf{R}_{b} = (0,0,+L/2)$;
$L$ is the superconductor thickness.
Importantly, one of these is a ``kink hedgehog''
while the other is an ``anti-kink hedgehog.''
Equations (\ref{eq:MajoranaKink})
and (\ref{eq:Esplit}) lead to:
\begin{eqnarray}
	E_{\text{split}}^{(z)}(L)
	&=&
		E_{0}^{(z)}\text{sech}^2\left( L/\xi_{so} \right) \\
	&\approx& 10^{-37} \text{ eV}
\end{eqnarray}
where the prefactor is
\begin{eqnarray}
	E_{0}^{(z)}
		&\equiv&
		-\frac{ 4 \sqrt{2} \Delta_{so}
			 \left[ K(-\epsilon_F^2 /\Delta^2) -\frac{E(-\epsilon_F^2 /\Delta^2)}{(1+\epsilon_F^2 /\Delta^2)} \right] }
			{\left[ K(-\epsilon_F^2 /\Delta^2) - E(-\epsilon_F^2 /\Delta^2) \right] }
\end{eqnarray}
While $E_{0}^{(z)}$ is seemingly a large energy scale, the sech$^2$ factor stemming from the in-plane localization
makes the degeneracy splitting due to coupling in the $z$-direction completely negligible compared to what we
calculate next.

For the lateral coupling within the same interface, we consider two Majoranas
$\Psi_{a}$ at $\mathbf{R}_{a} = (-R/2,0,-L/2)$ and
$\Psi_{b}$ at $\mathbf{R}_{b} = (+R/2,0,-L/2)$;
$R$ is the in-plane distance between hedgehogs.
Unlike the case of vertical coupling, the two wavefunctions here are both kinks.
Equations (\ref{eq:MajoranaKink}) and (\ref{eq:Esplit}) lead to:
\begin{eqnarray}
	E_{\text{split}}^{(xy)} (R,L)
	&=& \frac{E_{0}^{(xy)}(L) \cos\left( \epsilon_F R / \hbar v_F+ \alpha \right)}{\sqrt{R/\xi}} e^{-\frac{R}{\pi\xi}}
\end{eqnarray}
where the  $L$-dependent prefactor is
\begin{eqnarray}
	E_{0}^{(xy)} (L)
		&\equiv&
			\frac
			{8 \epsilon_F\tanh(L/\xi_{so}) [1+(\epsilon_F  / \Delta)^2 ]^{-1/4} }
			{ [E(-\epsilon_F^2 /\Delta^2) - K(-\epsilon_F^2 /\Delta^2) ]} \\
	E_{0}^{(xy)} (1000\text{\AA})
		&\approx&
			10.3 \text{ meV}
\end{eqnarray}
and the phase shift is $\alpha \equiv (1/2)\arctan \epsilon_F /\Delta$.
At the fields of interest to us, $E_{\text{split}}^{(xy)}(R,L)$ is much larger than $E_{\text{split}}^{(z)}(L)$,
so we neglect the latter.
The attenuation and period of oscillation of $E_{\text{split}}^{(xy)}(R,L)$ depends on the separation
between Majoranas, $R$, which is in turn determined by the lattice spacing of the Abrikosov
vortex lattice.  For a triangular lattice, this spacing is related to the field by
$R = \sqrt{\frac{h/2e}{B\sqrt{3}/2}}$.
Thus, the envelope of the energy splitting varies with field as
$E_{\text{split}}^{(xy)} (B) \propto \sqrt{B/B_0}e^{-\sqrt{B_0/B}}$,
where $B_0 \equiv \frac{2h/2e}{\sqrt{3} \pi^2 \xi^2} \approx 1.25$ Tesla.
This sets the lower bound of the temperature window and is clearly field-dependent:
$T_{L}(B) = E^{xy}_{\text{split}}(B)/k_B$.  See Fig.~\ref{fig:tempWindow}a.
We next determine the upper bound of the temperature window.

\section{Entropy}
\label{sec:Entropy}
To compute the cooling effect we need to understand all appreciable contributions to the total entropy as
a function of temperature and field: $S(T,B)$.  These can be classified into
phonon ($S_{ph}$)
and vortex ($S_{v}$)
contributions with the latter being composed of several pieces
(electronic contributions at the temperatures of interest  are negligible because $\Delta, \Delta_{so} \gg k_B T$).
Thus, the total entropy is given by
\begin{eqnarray}
S(T,B) = S_{ph}(T)+S_{v}(T,B)
\end{eqnarray}
The phonon entropy is standard:
\begin{equation}
S_{ph}(T) = k_B \frac{4\pi^4 V}{5V_{uc}}\left(\frac{T}{\Theta_D} \right)^3
\end{equation}
where $\Theta_D \approx 182$ K is the Debye temperature of the parent compound \cite{Shoemake1969} and
$V_{uc} \approx 426 \text{ \AA}^3$ is the volume of the unit cell \cite{Hor2010}.
Within the superconductor material Cu$_x$Bi$_2$Se$_3$, recent specific heat data \cite{Kriener2011} has yielded a Debye
temperature of $\Theta_D \approx 120$ K.

The vortex entropy will have several contributions.
The most important piece of $S_{v}$ is due to
the non-Abelian anyons ($S_{na}$); this is what drives the dramatic low-temperature cooling effect.
For a large number of vortices ($N_v$) this is simply \cite{Gervais2010}
\begin{equation}
	S_{na}(B) \approx N_v(B) 2k_B\log{\sqrt{2}}.
\end{equation}
The non-Abelian anyon entropy only depends on $B$, not $T$, and this is through its dependence on the number of vortices:
$S_{na}(B) \propto N_v(B)$ with $N_v(B) \approx BA/\phi_0$ where $\phi_0 = h/2e$ is the flux quantum and $A$ is the sample area
perpendicular to the magnetic field.
This linear approximation is justified for $B \gg B_{c1}$.

The contribution to the vortex entropy from more conventional sources will depend on the vortex density, and thus the magnetic field.
An isolated vortex line will contribute two types of entropy.
First, there are subgap bound states that are localized to the core but extended along the $z$-direction.
These are usually called CdGM excitations after Caroli, de Gennes, and Matricon \cite{Caroli1964}.
Second, a single vortex line will have excitations analogous to those of a fluctuating string \cite{Fetter1967}.
For our materials and parameter regimes, this second type of fluctuation turns out to contribute negligibly
to the total entropy.

In addition to this single-vortex physics, collective effects can manifest when the magnetic field is increased even slightly
above $B_{c1}$ leading to the formation of an Abrikosov vortex lattice.
A dense array of vortices can have collective modes of the same two types
as described above.  First, there are collective CdGM excitations \cite{Canel1965}, and second there are modes corresponding
to fluctuating elastic media \cite{Fetter1967}.  Note, however, that the term ``dense'' must be understood with respect to the appropriate length scale.
Define $R$ as the lateral vortex-vortex distance, $\lambda$ as the penetration depth, and $\xi$ as the
superconducting coherence length.
Collective modes of the vortex lattice appear when the magnetic field is such that $R < \lambda$.
In contrast, collective CdGM excitations only appear at the much higher density $R \sim \xi$.
For our device, $B_{c1} \approx 1.3$ mT
and $B_{c2} = 1.7$ T, giving a rather large range
$\xi \ll R \ll \lambda$
(or, equivalently, $B_{c1} \ll B \ll B_{c2}$)
in the dilute limit with respect to CdGM excitations,
but the dense collective-mode limit of vortex fluctuations.

In such a regime, the CdGM entropy is given by $N_{v}$ times the ``isolated'' vortex line contribution,
\begin{eqnarray}
	S_{\text{CdGM}}(T,B) &=& \frac{N_v(B) k_B 2k_F L}{\sqrt{2\pi} } \sqrt{\frac{g}{k_BT}}e^{-g/k_BT}
	\quad
\end{eqnarray}
while the vortex lattice entropy takes the form \cite{Fetter1967}
\begin{eqnarray}
	S_{vl}(T,B) &=& k_B\frac{15}{8}\zeta(5/2)\left(\frac{k_BT}{\hbar \bar{\kappa}V^{-2/3}} \right)^{3/2} \left(\frac{R(B)}{\lambda} \right)^{5/2}
	\quad
\end{eqnarray}
where
$g \equiv \frac{\Delta}{2k_F \xi}$ is the CdGM mini-gap,
$\bar{\kappa} \equiv h/2m \approx 2\times 10^{17} \text{\AA}^2/s$ is the quantum of circulation,
$\zeta$ is the Riemann-Zeta function,
and
$V = LA $ is the superconductor sample volume.
This form of $S_{vl}$, which is proportional to $B^{-5/4}$,
is only valid in the parameter regimes of interest to us.  Eventually, at very low fields on the order
of $B_{c1}$, $S_{vl}$ must of course vanish as $B$ decreases \cite{Fetter1967}.

Thus, the total vortex entropy is given by 
\begin{equation}
	S_v(T,B) = S_{na}(B)+S_{\text{CdGM}}(T,B)+S_{vl}(T,B)
\end{equation}
When added to the phonon entropy, this yields an approximate analytic expression for the
total system entropy as a function of $T$ and $B$.  Using these expressions we calculate
the central quantity $\frac{dT}{dB} = -\frac{(dS/dB)_T}{(dS/dT)_B}$ as described in the next section.

\section{Cooling}
\label{sec:Cooling}
To understand the cooling effect, consider the small change in entropy for a system depending
on temperature and field:
\begin{eqnarray}
	dS = \left(\frac{dS}{dT} \right)_{B}dT +  \left(\frac{dS}{dB} \right)_{T}dB
\end{eqnarray}
For an isentropic process ($dS=0$), the system's temperature changes in response to a small field change according to:
\begin{eqnarray}
	\frac{dT}{dB} = -\frac{(dS/dB)_T}{(dS/dT)_B}.
\end{eqnarray}
When this quantity is negative it represents a decrease of system's temperature as the field
is adiabatically increased: cooling.  In contrast, a positive sign indicates heating.
Importantly, the sign and magnitude of $\frac{dT}{dB}$ depend on $T$ and $B$.
Since $\frac{dS}{dT}$ is always positive, to find cooling we require a parameter regime in which $\frac{dS}{dB}$ is positive.
This will occur when $S_{na}(B)$ dominates the total entropy.

\begin{figure}[tbp]
   \centering
   \includegraphics[width=3.4in]{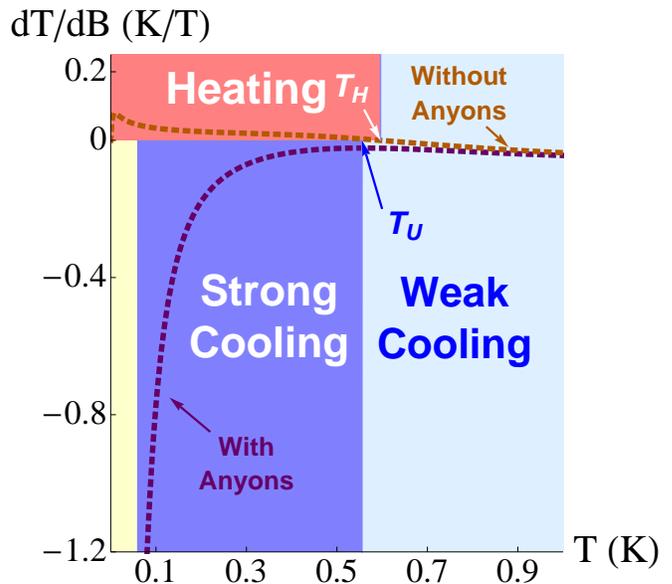}
   \caption{
   (color online) Quantitative illustration of cooling via 3D non-Abelian anyons. 
   The superconductor layer thickness is set to $L = 1000$ \AA .
   	The orange and purple dashed lines are
   	constant field cuts ($B = 25 B_{c1} \approx 33$ mT)  of $\frac{dT}{dB}$
   	corresponding to the orange and purple dashed lines in Fig.~\ref{fig:tempWindow} with (a) and without (b) anyons.
	Notice that $T_H \gtrapprox T_U$, where $T_H$ is defined by the temperature below which
	$\frac{dT}{dB}>0$ in a system without 3D Majoranas, and $T_U$, the upper bound of the cooling window,
	is defined as the temperature below which the non-Abelian entropy strongly enhances the cooling effect
	in a system with 3D Majoranas.
   }
   \label{fig:dTdB}
\end{figure}

In Fig.~\ref{fig:dTdB} we show the temperature dependence, both with and without 3D anyons in the system,
for a fixed value of field: $B = 25 B_{c1} \gg B_{c1}$.
At high temperatures $dT/dB$ is very similar in both cases, but below a certain temperature, $T_H$,
the system without 3D anyons will experience heating while the system with 3D anyons will experience cooling.
This qualitative difference is the harbinger of 3D non-Abelian anyons.

\begin{figure}[tbp]
   \centering
   \includegraphics[width=3.4in]{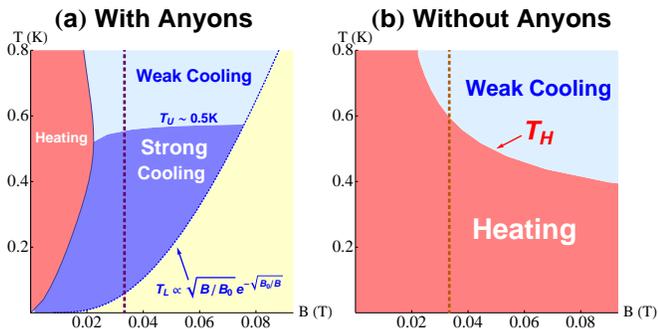}
   \caption{
   (color online) Comparison of the temperature evolution with (a) and without (b) 3D non-Abelian anyons
   in the field-temperature plane.
   The superconductor layer thickness is set to $L = 1000$ \AA .
   	The sign and magnitude of $\frac{dT}{dB}(B,T)$ determines the nature of the cooling or heating effect as
   	the magnetic field is adiabatically increased.  When 3D Majoranas are present (a), a sizable temperature window
	$T_L < T < T_U$ at intermediate fields (darker blue region labelled ``Strong Cooling'') will exhibit a strong cooling effect.
	Without 3D Majoranas (b), the same parameter regime will demonstrate a heating effect.
	The yellow region below $T_L(B)$ is where the ground state degeneracy is split by Majorana-Majorana tunneling; oscillations
	are not shown for clarity.
	This interesting sector may include collective anyon excitations generalizing some ideas in 2D \cite{Ludwig2011},
	but these are not pertinent to this paper.   
	}
   \label{fig:tempWindow}
\end{figure}

To reiterate: weak cooling may occur at high temperatures with or without 3D anyons,
but in a specific region of the $BT$-plane, depicted in dark blue in Fig~\ref{fig:tempWindow}a,
a system with 3D anyons will exhibit strong cooling while a system without 3D anyons will experience heating
(as shown in Fig~\ref{fig:tempWindow}b).  For a system with 3D anyons, the transition between the strong cooling
and weak cooling regimes is defined by the extremum of the curve $dT/dB$ which is most apparent by examining
the purple dashed curve in Fig.~\ref{fig:dTdB}.

To understand the origin of the minor heating without anyons note that,
in the regimes relevant to our proposal,
all contributions to the entropy are non-decreasing functions of $B$ \textit{except} $S_{vl}$.
This part of the entropy \textit{decreases} with field and thus $dS/dB$ could become negative if $S_{vl}$
were large enough compared to all the other entropy sources.
Since $dS/dT$ is always positive, the possibility of negative $dS/dB$ means $dT/dB = -\frac{dS/dB}{dS/dT}$
could be driven positive leading to the observation of heating with an adiabatic increase of the field.
Indeed, in the absence of 3D non-abelian anyons this is exactly what happens at low temperatures
as depicted in Fig.~\ref{fig:tempWindow}b.  However, the presence of 3D non-abelian anyons
contributes $S_{na}$ to the total entropy which increases sufficiently rapidly with $B$ that it overwhelms
the low temperature heating due to $S_{vl}$ and reverses the trend to produce a dramatic strong cooling effect.

\section{Conclusion}
\label{sec:Conclusions}
In summary, we have quantitatively estimated a magnetic field and temperature regime in which an experiment
might detect 3D non-Abelian anyons in a heterostructure device composed of Bi$_2$Se$_3$ and Cu$_{0.12}$Bi$_2$Se$_3$.
Within this regime, a system with 3D non-Abelian anyons will experience a decrease in temperature as the magnetic field is adiabatically increased. In contrast, a system without 3D non-Abelian anyons under identical conditions will exhibit a temperature increase
as the magnetic field is adiabatically increased.  In the strong cooling region, the effect is rather large.  For example,
at $B = 25 B_{c1}$ and $T=0.1$ K we have $\frac{dT}{dB} \approx -1.7 $ K/T, which should be detectable with
present technology.

We close by enumerating a few caveats that are specific to Cu$_{0.12}$Bi$_2$Se$_3$, 
but not to the general idea of our proposal.
First, the numbers presented in this paper are based on a
value of the superconducting gap inferred from the $B_{c2}$-derived coherence length \cite{Hor2010}
which disagrees with the $T_c$ derived gap \cite{Wray2010}.
Since the degeneracy splitting (\textit{i.e.} the $T_L(B)$ line) is exponentially sensitive to this parameter, the choice
is very important.  We hope a direct experimental measurement of the superconducting gap will be made in the future
to pin down the true value of this important parameter.
Second, we have assumed the sign of $\Delta_{so}$ to take opposite values in insulating Bi$_2$Se$_3$
versus Cu$_{0.12}$Bi$_2$Se$_3$.  Without the sign change, there will be no Majorana mode.
Again, this needs to be experimentally checked for Cu$_{0.12}$Bi$_2$Se$_3$.
Note, however, that if later experimental investigations determine that either $\Delta$ in Cu$_{0.12}$Bi$_2$Se$_3$ is much smaller
than our assumption, or that $\Delta_{so}$ does not change sign in Cu$_{0.12}$Bi$_2$Se$_3$, the general
idea of this proposal will not be invalidated but only its applicability to this particular superconductor;
all qualitative conclusions will remain true for \textit{any} insulator-superconductor system that satisfies the following conditions:
(i) the superconductor must have a relatively large s-wave gap and be strongly type II;
(ii) the band gap must take opposite signs in the insulating and superconducting regions.
These are relatively simple conditions to fulfill.  We chose to examine a Bi$_2$Se$_3$-based structure because this material
has emerged as an archtype topological insulator which is being independently studied by many different research institutions.
Furthermore, the possibility of creating topological insulator regions and superconducting regions simply by p-doping or n-doping 
the same parent compound makes Bi$_2$Se$_3$ a very attractive system for building heterostructure devices in the future.

\section{Acknowledgements }
This work is supported by DOE under Grant No.~DE-FG52-10NA29659 (SJY) and NSF Grant No. DMR-1004545 (SJY and KY).


\begin{thebibliography}{39}
\expandafter\ifx\csname natexlab\endcsname\relax\def\natexlab#1{#1}\fi
\expandafter\ifx\csname bibnamefont\endcsname\relax
  \def\bibnamefont#1{#1}\fi
\expandafter\ifx\csname bibfnamefont\endcsname\relax
  \def\bibfnamefont#1{#1}\fi
\expandafter\ifx\csname citenamefont\endcsname\relax
  \def\citenamefont#1{#1}\fi
\expandafter\ifx\csname url\endcsname\relax
  \def\url#1{\texttt{#1}}\fi
\expandafter\ifx\csname urlprefix\endcsname\relax\def\urlprefix{URL }\fi
\providecommand{\bibinfo}[2]{#2}
\providecommand{\eprint}[2][]{\url{#2}}

\bibitem[{\citenamefont{{N}ayak \textit{et al.}}(2008)}]{Nayak2008}
\bibinfo{author}{\bibfnamefont{C.}~\bibnamefont{{N}ayak \textit{et al.}}},
  \bibinfo{journal}{Rev. Mod. Phys.} \textbf{\bibinfo{volume}{80}},
  \bibinfo{pages}{1083} (\bibinfo{year}{2008}).

\bibitem[{\citenamefont{{Teo} and {Kane}}(2010)}]{Teo2010}
\bibinfo{author}{\bibfnamefont{J.~C.~Y.} \bibnamefont{{Teo}}} \bibnamefont{and}
  \bibinfo{author}{\bibfnamefont{C.~L.} \bibnamefont{{Kane}}},
  \bibinfo{journal}{\prl} \textbf{\bibinfo{volume}{104}},
  \bibinfo{pages}{046401} (\bibinfo{year}{2010}).

\bibitem[{\citenamefont{{Freedman} et~al.}(2011)\citenamefont{{Freedman},
  {Hastings}, {Nayak}, {Qi}, {Walker}, and {Wang}}}]{Freedman2011}
\bibinfo{author}{\bibfnamefont{M.}~\bibnamefont{{Freedman}}},
  \bibinfo{author}{\bibfnamefont{M.~B.} \bibnamefont{{Hastings}}},
  \bibinfo{author}{\bibfnamefont{C.}~\bibnamefont{{Nayak}}},
  \bibinfo{author}{\bibfnamefont{X.~L.} \bibnamefont{{Qi}}},
  \bibinfo{author}{\bibfnamefont{K.}~\bibnamefont{{Walker}}}, \bibnamefont{and}
  \bibinfo{author}{\bibfnamefont{Z.}~\bibnamefont{{Wang}}},
  \bibinfo{journal}{\prb} \textbf{\bibinfo{volume}{83}},
  \bibinfo{pages}{115132} (\bibinfo{year}{2011}).

\bibitem[{\citenamefont{{Yang} and {Halperin}}(2009)}]{Yang2009}
\bibinfo{author}{\bibfnamefont{K.}~\bibnamefont{{Yang}}} \bibnamefont{and}
  \bibinfo{author}{\bibfnamefont{B.~I.} \bibnamefont{{Halperin}}},
  \bibinfo{journal}{\prb} \textbf{\bibinfo{volume}{79}},
  \bibinfo{pages}{115317} (\bibinfo{year}{2009}).

\bibitem[{\citenamefont{{Cooper} and {Stern}}(2009)}]{Cooper2009}
\bibinfo{author}{\bibfnamefont{N.~R.} \bibnamefont{{Cooper}}} \bibnamefont{and}
  \bibinfo{author}{\bibfnamefont{A.}~\bibnamefont{{Stern}}},
  \bibinfo{journal}{\prl} \textbf{\bibinfo{volume}{102}},
  \bibinfo{pages}{176807} (\bibinfo{year}{2009}).

\bibitem[{\citenamefont{{Gervais} and {Yang}}(2010)}]{Gervais2010}
\bibinfo{author}{\bibfnamefont{G.}~\bibnamefont{{Gervais}}} \bibnamefont{and}
  \bibinfo{author}{\bibfnamefont{K.}~\bibnamefont{{Yang}}},
  \bibinfo{journal}{\prl} \textbf{\bibinfo{volume}{105}},
  \bibinfo{pages}{086801} (\bibinfo{year}{2010}).

\bibitem[{\citenamefont{{Stern} and {Halperin}}(2006)}]{Stern2006}
\bibinfo{author}{\bibfnamefont{A.}~\bibnamefont{{Stern}}} \bibnamefont{and}
  \bibinfo{author}{\bibfnamefont{B.~I.} \bibnamefont{{Halperin}}},
  \bibinfo{journal}{\prl} \textbf{\bibinfo{volume}{96}},
  \bibinfo{pages}{016802} (\bibinfo{year}{2006}).

\bibitem[{\citenamefont{{Bonderson} et~al.}(2006)\citenamefont{{Bonderson},
  {Kitaev}, and {Shtengel}}}]{Bonderson2006}
\bibinfo{author}{\bibfnamefont{P.}~\bibnamefont{{Bonderson}}},
  \bibinfo{author}{\bibfnamefont{A.}~\bibnamefont{{Kitaev}}}, \bibnamefont{and}
  \bibinfo{author}{\bibfnamefont{K.}~\bibnamefont{{Shtengel}}},
  \bibinfo{journal}{\prl} \textbf{\bibinfo{volume}{96}},
  \bibinfo{pages}{016803} (\bibinfo{year}{2006}).

\bibitem[{\citenamefont{{Willett} et~al.}(2009)\citenamefont{{Willett},
  {Pfeiffer}, and {West}}}]{Willett2009}
\bibinfo{author}{\bibfnamefont{R.~L.} \bibnamefont{{Willett}}},
  \bibinfo{author}{\bibfnamefont{L.~N.} \bibnamefont{{Pfeiffer}}},
  \bibnamefont{and} \bibinfo{author}{\bibfnamefont{K.~W.}
  \bibnamefont{{West}}}, \bibinfo{journal}{Proc. Natl. Acad. Sci.}
  \textbf{\bibinfo{volume}{106}}, \bibinfo{pages}{8853} (\bibinfo{year}{2009}).

\bibitem[{\citenamefont{{Willett} et~al.}(2010)\citenamefont{{Willett},
  {Pfeiffer}, and {West}}}]{Willett2010}
\bibinfo{author}{\bibfnamefont{R.~L.} \bibnamefont{{Willett}}},
  \bibinfo{author}{\bibfnamefont{L.~N.} \bibnamefont{{Pfeiffer}}},
  \bibnamefont{and} \bibinfo{author}{\bibfnamefont{K.~W.}
  \bibnamefont{{West}}}, \bibinfo{journal}{\prb} \textbf{\bibinfo{volume}{82}},
  \bibinfo{pages}{205301} (\bibinfo{year}{2010}).

\bibitem[{\citenamefont{{Das~Sarma} et~al.}(2006)\citenamefont{{Das~Sarma},
  {Nayak}, and {Tewari}}}]{DasSarma2006}
\bibinfo{author}{\bibfnamefont{S.}~\bibnamefont{{Das~Sarma}}},
  \bibinfo{author}{\bibfnamefont{C.}~\bibnamefont{{Nayak}}}, \bibnamefont{and}
  \bibinfo{author}{\bibfnamefont{S.}~\bibnamefont{{Tewari}}},
  \bibinfo{journal}{\prb} \textbf{\bibinfo{volume}{73}},
  \bibinfo{pages}{220502} (\bibinfo{year}{2006}).

\bibitem[{\citenamefont{{Tsutsumi} et~al.}(2008)\citenamefont{{Tsutsumi},
  {Kawakami}, {Mizushima}, {Ichioka}, and {Machida}}}]{Tsutsumi2008}
\bibinfo{author}{\bibfnamefont{Y.}~\bibnamefont{{Tsutsumi}}},
  \bibinfo{author}{\bibfnamefont{T.}~\bibnamefont{{Kawakami}}},
  \bibinfo{author}{\bibfnamefont{T.}~\bibnamefont{{Mizushima}}},
  \bibinfo{author}{\bibfnamefont{M.}~\bibnamefont{{Ichioka}}},
  \bibnamefont{and}
  \bibinfo{author}{\bibfnamefont{K.}~\bibnamefont{{Machida}}},
  \bibinfo{journal}{\prl} \textbf{\bibinfo{volume}{101}},
  \bibinfo{pages}{135302} (\bibinfo{year}{2008}).

\bibitem[{\citenamefont{{Fu} and {Kane}}(2009)}]{Fu2009}
\bibinfo{author}{\bibfnamefont{L.}~\bibnamefont{{Fu}}} \bibnamefont{and}
  \bibinfo{author}{\bibfnamefont{C.~L.} \bibnamefont{{Kane}}},
  \bibinfo{journal}{\prl} \textbf{\bibinfo{volume}{102}},
  \bibinfo{pages}{216403} (\bibinfo{year}{2009}).

\bibitem[{\citenamefont{{Sau} et~al.}(2010)\citenamefont{{Sau}, {Lutchyn},
  {Tewari}, and {Das Sarma}}}]{Sau2010}
\bibinfo{author}{\bibfnamefont{J.~D.} \bibnamefont{{Sau}}},
  \bibinfo{author}{\bibfnamefont{R.~M.} \bibnamefont{{Lutchyn}}},
  \bibinfo{author}{\bibfnamefont{S.}~\bibnamefont{{Tewari}}}, \bibnamefont{and}
  \bibinfo{author}{\bibfnamefont{S.}~\bibnamefont{{Das Sarma}}},
  \bibinfo{journal}{\prl} \textbf{\bibinfo{volume}{104}},
  \bibinfo{pages}{040502} (\bibinfo{year}{2010}).

\bibitem[{\citenamefont{{Alicea}}(2010)}]{Alicea2010}
\bibinfo{author}{\bibfnamefont{J.}~\bibnamefont{{Alicea}}},
  \bibinfo{journal}{\prb} \textbf{\bibinfo{volume}{81}},
  \bibinfo{pages}{125318} (\bibinfo{year}{2010}).

\bibitem[{\citenamefont{{Zhu} et~al.}(2011)\citenamefont{{Zhu}, {Shao}, {Wang},
  and {Duan}}}]{Zhu2011}
\bibinfo{author}{\bibfnamefont{S.~L.} \bibnamefont{{Zhu}}},
  \bibinfo{author}{\bibfnamefont{L.~B.} \bibnamefont{{Shao}}},
  \bibinfo{author}{\bibfnamefont{Z.~D.} \bibnamefont{{Wang}}},
  \bibnamefont{and} \bibinfo{author}{\bibfnamefont{L.~M.}
  \bibnamefont{{Duan}}}, \bibinfo{journal}{\prl}
  \textbf{\bibinfo{volume}{106}}, \bibinfo{pages}{100404}
  (\bibinfo{year}{2011}).

\bibitem[{\citenamefont{Kitaev}(2001)}]{Kitaev2001}
\bibinfo{author}{\bibfnamefont{A.~Y.} \bibnamefont{Kitaev}},
  \bibinfo{journal}{Phys.-Usp} \textbf{\bibinfo{volume}{44}},
  \bibinfo{pages}{131} (\bibinfo{year}{2001}).

\bibitem[{\citenamefont{{Oreg} et~al.}(2010)\citenamefont{{Oreg}, {Refael}, and
  {von Oppen}}}]{Oreg2010}
\bibinfo{author}{\bibfnamefont{Y.}~\bibnamefont{{Oreg}}},
  \bibinfo{author}{\bibfnamefont{G.}~\bibnamefont{{Refael}}}, \bibnamefont{and}
  \bibinfo{author}{\bibfnamefont{F.}~\bibnamefont{{von Oppen}}},
  \bibinfo{journal}{\prl} \textbf{\bibinfo{volume}{105}},
  \bibinfo{pages}{177002} (\bibinfo{year}{2010}).

\bibitem[{\citenamefont{{Lutchyn} et~al.}(2010)\citenamefont{{Lutchyn}, {Sau},
  and {Das Sarma}}}]{Lutchyn2010}
\bibinfo{author}{\bibfnamefont{R.~M.} \bibnamefont{{Lutchyn}}},
  \bibinfo{author}{\bibfnamefont{J.~D.} \bibnamefont{{Sau}}}, \bibnamefont{and}
  \bibinfo{author}{\bibfnamefont{S.}~\bibnamefont{{Das Sarma}}},
  \bibinfo{journal}{\prl} \textbf{\bibinfo{volume}{105}},
  \bibinfo{pages}{077001} (\bibinfo{year}{2010}).

\bibitem[{\citenamefont{{Alicea} et~al.}(2011)\citenamefont{{Alicea}, {Oreg},
  {Refael}, {von Oppen}, and {Fisher}}}]{Alicea2011}
\bibinfo{author}{\bibfnamefont{J.}~\bibnamefont{{Alicea}}},
  \bibinfo{author}{\bibfnamefont{Y.}~\bibnamefont{{Oreg}}},
  \bibinfo{author}{\bibfnamefont{G.}~\bibnamefont{{Refael}}},
  \bibinfo{author}{\bibfnamefont{F.}~\bibnamefont{{von Oppen}}},
  \bibnamefont{and} \bibinfo{author}{\bibfnamefont{M.~P.~A.}
  \bibnamefont{{Fisher}}}, \bibinfo{journal}{Nature Physics}
  \textbf{\bibinfo{volume}{7}}, \bibinfo{pages}{412} (\bibinfo{year}{2011}).

\bibitem[{\citenamefont{{Nayak}}(2010)}]{Nayak2010}
\bibinfo{author}{\bibfnamefont{C.}~\bibnamefont{{Nayak}}},
  \bibinfo{journal}{\nat} \textbf{\bibinfo{volume}{464}}, \bibinfo{pages}{693}
  (\bibinfo{year}{2010}).

\bibitem[{\citenamefont{{Stern} and {Levin}}(2010)}]{Stern2010}
\bibinfo{author}{\bibfnamefont{A.}~\bibnamefont{{Stern}}} \bibnamefont{and}
  \bibinfo{author}{\bibfnamefont{M.}~\bibnamefont{{Levin}}},
  \bibinfo{journal}{Physics Online Journal} \textbf{\bibinfo{volume}{3}},
  \bibinfo{pages}{7} (\bibinfo{year}{2010}).

\bibitem[{\citenamefont{{H}or \textit{et al.}}(2010)}]{Hor2010}
\bibinfo{author}{\bibfnamefont{Y.~S.} \bibnamefont{{H}or \textit{et al.}}},
  \bibinfo{journal}{\prl} \textbf{\bibinfo{volume}{104}},
  \bibinfo{pages}{057001} (\bibinfo{year}{2010}).

\bibitem[{\citenamefont{{Fu} and {Berg}}(2010)}]{Fu2010}
\bibinfo{author}{\bibfnamefont{L.}~\bibnamefont{{Fu}}} \bibnamefont{and}
  \bibinfo{author}{\bibfnamefont{E.}~\bibnamefont{{Berg}}},
  \bibinfo{journal}{\prl} \textbf{\bibinfo{volume}{105}},
  \bibinfo{pages}{097001} (\bibinfo{year}{2010}).

\bibitem[{\citenamefont{{W}ray \textit{et al.}}(2010)}]{Wray2010}
\bibinfo{author}{\bibfnamefont{L.~A.} \bibnamefont{{W}ray \textit{et al.}}},
  \bibinfo{journal}{Nat. Phys.} \textbf{\bibinfo{volume}{6}},
  \bibinfo{pages}{855} (\bibinfo{year}{2010}).

\bibitem[{\citenamefont{{Kriener} et~al.}(2011)\citenamefont{{Kriener},
  {Segawa}, {Ren}, {Sasaki}, and {Ando}}}]{Kriener2011}
\bibinfo{author}{\bibfnamefont{M.}~\bibnamefont{{Kriener}}},
  \bibinfo{author}{\bibfnamefont{K.}~\bibnamefont{{Segawa}}},
  \bibinfo{author}{\bibfnamefont{Z.}~\bibnamefont{{Ren}}},
  \bibinfo{author}{\bibfnamefont{S.}~\bibnamefont{{Sasaki}}}, \bibnamefont{and}
  \bibinfo{author}{\bibfnamefont{Y.}~\bibnamefont{{Ando}}},
  \bibinfo{journal}{Phys. Rev. Lett.} \textbf{\bibinfo{volume}{106}},
  \bibinfo{pages}{127004} (\bibinfo{year}{2011}).

\bibitem[{\citenamefont{{Nishida} et~al.}(2010)\citenamefont{{Nishida},
  {Santos}, and {Chamon}}}]{Nishida2010}
\bibinfo{author}{\bibfnamefont{Y.}~\bibnamefont{{Nishida}}},
  \bibinfo{author}{\bibfnamefont{L.}~\bibnamefont{{Santos}}}, \bibnamefont{and}
  \bibinfo{author}{\bibfnamefont{C.}~\bibnamefont{{Chamon}}},
  \bibinfo{journal}{\prb} \textbf{\bibinfo{volume}{82}},
  \bibinfo{pages}{144513} (\bibinfo{year}{2010}).

\bibitem[{\citenamefont{{Fukui}}(2010)}]{Fukui2010}
\bibinfo{author}{\bibfnamefont{T.}~\bibnamefont{{Fukui}}},
  \bibinfo{journal}{\prb} \textbf{\bibinfo{volume}{81}},
  \bibinfo{pages}{214516} (\bibinfo{year}{2010}).

\bibitem[{\citenamefont{{Jackiw} and {Rebbi}}(1976)}]{Jackiw1976}
\bibinfo{author}{\bibfnamefont{R.}~\bibnamefont{{Jackiw}}} \bibnamefont{and}
  \bibinfo{author}{\bibfnamefont{C.}~\bibnamefont{{Rebbi}}},
  \bibinfo{journal}{\prd} \textbf{\bibinfo{volume}{13}}, \bibinfo{pages}{3398}
  (\bibinfo{year}{1976}).

\bibitem[{\citenamefont{{Liu} et~al.}(2010)\citenamefont{{Liu}, {Qi}, {Zhang},
  {Dai}, {Fang}, and {Zhang}}}]{Liu2010}
\bibinfo{author}{\bibfnamefont{C.-X.} \bibnamefont{{Liu}}},
  \bibinfo{author}{\bibfnamefont{X.-L.} \bibnamefont{{Qi}}},
  \bibinfo{author}{\bibfnamefont{H.~J.} \bibnamefont{{Zhang}}},
  \bibinfo{author}{\bibfnamefont{X.}~\bibnamefont{{Dai}}},
  \bibinfo{author}{\bibfnamefont{Z.}~\bibnamefont{{Fang}}}, \bibnamefont{and}
  \bibinfo{author}{\bibfnamefont{S.-C.} \bibnamefont{{Zhang}}},
  \bibinfo{journal}{\prb} \textbf{\bibinfo{volume}{82}},
  \bibinfo{pages}{045122} (\bibinfo{year}{2010}).

\bibitem[{\citenamefont{{Gygi} and {Schl{\"u}ter}}(1991)}]{Gygi1991}
\bibinfo{author}{\bibfnamefont{F.}~\bibnamefont{{Gygi}}} \bibnamefont{and}
  \bibinfo{author}{\bibfnamefont{M.}~\bibnamefont{{Schl{\"u}ter}}},
  \bibinfo{journal}{\prb} \textbf{\bibinfo{volume}{43}}, \bibinfo{pages}{7609}
  (\bibinfo{year}{1991}).

\bibitem[{\citenamefont{{Cheng} et~al.}(2010)\citenamefont{{Cheng}, {Lutchyn},
  {Galitski}, and {Das Sarma}}}]{Cheng2010}
\bibinfo{author}{\bibfnamefont{M.}~\bibnamefont{{Cheng}}},
  \bibinfo{author}{\bibfnamefont{R.~M.} \bibnamefont{{Lutchyn}}},
  \bibinfo{author}{\bibfnamefont{V.}~\bibnamefont{{Galitski}}},
  \bibnamefont{and} \bibinfo{author}{\bibfnamefont{S.}~\bibnamefont{{Das
  Sarma}}}, \bibinfo{journal}{\prb} \textbf{\bibinfo{volume}{82}},
  \bibinfo{pages}{094504} (\bibinfo{year}{2010}).

\bibitem[{\citenamefont{{Landau} and {Lifshits}}(1974)}]{Landau1974}
\bibinfo{author}{\bibfnamefont{L.~D.} \bibnamefont{{Landau}}} \bibnamefont{and}
  \bibinfo{author}{\bibfnamefont{E.~M.} \bibnamefont{{Lifshits}}},
  \emph{\bibinfo{title}{{Quantum Mechanics. Nonrelativistic theory}}}
  (\bibinfo{publisher}{Pergamon Press}, \bibinfo{year}{1974}), pp.
  \bibinfo{pages}{183--184}, \bibinfo{edition}{3rd} ed.

\bibitem[{\citenamefont{{Mizushima} and {Machida}}(2010)}]{Mizushima2010}
\bibinfo{author}{\bibfnamefont{T.}~\bibnamefont{{Mizushima}}} \bibnamefont{and}
  \bibinfo{author}{\bibfnamefont{K.}~\bibnamefont{{Machida}}},
  \bibinfo{journal}{\pra} \textbf{\bibinfo{volume}{82}},
  \bibinfo{pages}{023624} (\bibinfo{year}{2010}).

\bibitem[{\citenamefont{{Shoemake} et~al.}(1969)\citenamefont{{Shoemake},
  {Rayne}, and {Ure}}}]{Shoemake1969}
\bibinfo{author}{\bibfnamefont{G.~E.} \bibnamefont{{Shoemake}}},
  \bibinfo{author}{\bibfnamefont{J.~A.} \bibnamefont{{Rayne}}},
  \bibnamefont{and} \bibinfo{author}{\bibfnamefont{R.~W.} \bibnamefont{{Ure}}},
  \bibinfo{journal}{Phys. Rev.} \textbf{\bibinfo{volume}{185}},
  \bibinfo{pages}{1046} (\bibinfo{year}{1969}).

\bibitem[{\citenamefont{{Caroli} et~al.}(1964)\citenamefont{{Caroli}, {de
  Gennes}, and {Matricon}}}]{Caroli1964}
\bibinfo{author}{\bibfnamefont{C.}~\bibnamefont{{Caroli}}},
  \bibinfo{author}{\bibfnamefont{P.~G.} \bibnamefont{{de Gennes}}},
  \bibnamefont{and}
  \bibinfo{author}{\bibfnamefont{J.}~\bibnamefont{{Matricon}}},
  \bibinfo{journal}{Phys. Lett.} \textbf{\bibinfo{volume}{9}},
  \bibinfo{pages}{307} (\bibinfo{year}{1964}).

\bibitem[{\citenamefont{{Fetter}}(1967)}]{Fetter1967}
\bibinfo{author}{\bibfnamefont{A.~L.} \bibnamefont{{Fetter}}},
  \bibinfo{journal}{Phys. Rev.} \textbf{\bibinfo{volume}{163}},
  \bibinfo{pages}{390} (\bibinfo{year}{1967}).

\bibitem[{\citenamefont{{Canel}}(1965)}]{Canel1965}
\bibinfo{author}{\bibfnamefont{E.}~\bibnamefont{{Canel}}},
  \bibinfo{journal}{Phys. Lett.} \textbf{\bibinfo{volume}{16}},
  \bibinfo{pages}{101} (\bibinfo{year}{1965}).

\bibitem[{\citenamefont{{Ludwig} et~al.}(2011)\citenamefont{{Ludwig},
  {Poilblanc}, {Trebst}, and {Troyer}}}]{Ludwig2011}
\bibinfo{author}{\bibfnamefont{A.~W.~W.} \bibnamefont{{Ludwig}}},
  \bibinfo{author}{\bibfnamefont{D.}~\bibnamefont{{Poilblanc}}},
  \bibinfo{author}{\bibfnamefont{S.}~\bibnamefont{{Trebst}}}, \bibnamefont{and}
  \bibinfo{author}{\bibfnamefont{M.}~\bibnamefont{{Troyer}}},
  \bibinfo{journal}{New J. of Phys.} \textbf{\bibinfo{volume}{13}},
  \bibinfo{pages}{045014} (\bibinfo{year}{2011}).

\end{thebibliography}

\end{document}